\documentclass{article}
\usepackage{spconf,amsmath,graphicx}

\usepackage{xcolor}
\usepackage{multirow}
\usepackage{float}
\usepackage{cite}
\newcommand{\jiatong}[1]{\textcolor{blue}{\bf\small [#1 --jiatong]}}

\usepackage{booktabs,multirow}
\usepackage{adjustbox}

\renewcommand{\vec}[1]{\mathbf{#1}}

\title{Improving RNN transducer with target speaker extraction and neural uncertainty estimation}
\name{Jiatong Shi$^{\star}$\sthanks{Jiatong Shi performed the work during his internship at Tencent AI Lab.}, Chunlei Zhang$^{\dagger}$, Chao Weng$^{\dagger}$, Shinji Watanabe$^{\star}$, Meng Yu$^{\dagger}$, Dong Yu$^{\dagger}$ }
\address{
    $^{\star}$ Johns Hopkins University, USA\\
    $^{\dagger}$ Tencent AI Lab, Bellevue WA, USA\\
    \small{\texttt{\{jiatong\_shi, shinjw\}@jhu.edu, \{cleizhang, cweng, raymondmyu, dyu\}@tencent.com}}
}

\begin{document}
\ninept
\maketitle
\begin{abstract}
Target-speaker speech recognition aims to recognize target-speaker speech from noisy environments with background noise and interfering speakers. 
This work presents a joint framework that combines time-domain target-speaker speech extraction and Recurrent Neural Network Transducer (RNN-T). 
To stabilize the joint-training, we propose a multi-stage training strategy that pre-trains and fine-tunes each module in the system before joint-training. 
Meanwhile, speaker identity and speech enhancement uncertainty measures are proposed to compensate for residual noise and artifacts from the target speech extraction module.
Compared to a recognizer fine-tuned with a target speech extraction model, our experiments show that adding the neural uncertainty module significantly reduces 17\% relative Character Error Rate (CER) on multi-speaker signals with background noise. The multi-condition experiments indicate that our method can achieve 9\% relative performance gain in the noisy condition while maintaining the performance in the clean condition.
\end{abstract}

\section{Introduction}

Target-speaker speech extraction is a process to separate specific speaker's speech from a speech mixture by eliminating other interfering speakers and noises. 
Conventional methods for the problem include anchored speech detection \cite{king2017robust, mallidi2018device} and speaker-specific training \cite{zhang2016deep, du2016regression}. However, both methods have their limitations when dealing with the problem. The first can only handle non-overlap speech, while the second cannot be applied to unseen speakers without re-training of the entire network. 
State-of-the-art methods adopt a couple of seconds of enrollment target-speaker speech to extract a target-speaker embedding and employ it as auxiliary information to enhancement and separation networks \cite{zmolikova2017speaker, delcroix2018single, vzmolikova2019speakerbeam, wang2019voicefilter, ji2020speaker, wang2020voicefilterlite}.

One of the applications of target-speaker speech extraction is to use it as a prepossessing module of Automatic Speech Recognition (ASR) systems. 
Previous literature investigated the target-speaker speech extraction with different back-end ASR architectures. 
In \cite{wang2019voicefilter}, Wang et al. adopted a word-based Connectionist temporal classification recognizer. HMM-DNN (Hidden Markov Model-Deep Neural Networks) was employed in \cite{zmolikova2017speaker, delcroix2018single, vzmolikova2019speakerbeam}. Wang et al. tested the Voicefilter-Lite with an RNN-T \cite{wang2020voicefilterlite}. 
Aforementioned works mainly investigated the Time-Frequency (TF) domain target speech extraction. \v{Z}mol\'ikov\'a et al. have studied the joint-training of target speaker extraction and HMM-based ASR \cite{delcroix2018single, vzmolikova2019speakerbeam}. Others, however, employed a pre-trained recognizer without joint-training. 

On the other hand, in the other robust ASR field including speech enhancement/separation, joint training becomes a popular methodology to make a system robust against the mismatch between the preprocessing and ASR modules \cite{narayanan2014joint, gao2015joint,  heymann2017beamnet, ochiai2017multichannel, settle2018end, subramanian2020far, xu2019joint}.
Similar to target speaker ASR, most previous studies investigated the joint-training on the TF domain.
However, in recent work, Neumann et al. showed that the joint-training with time-domain speech separation and encoder-decoder-based ASR could significantly elevate the performances on multi-talker speech recognition \cite{von2020multi}.


Given this trend, this paper proposes to subsume time-domain target-speaker speech extraction into an RNN-T end-to-end ASR system.
Practically, RNN-T end-to-end speech recognizer is a good candidate for on-device scenarios because of its small system footprints. 
Also, the integration of the target speaker extraction module in this on-device system can avoid the requirement for users to upload their enrollment voice, which greatly mitigates the concern of users' privacy.
%
However, one of the important problems for this joint training is how to compensate for residual noises and artifacts introduced from the extraction process.
To deal with this issue, we leverage a neural network based speech enhancement and speaker identity uncertainty estimation \cite{deng2005dynamic, tran2014fusion} for the joint system. 
This uncertainty information is then used as a feature for ASR training. 
Our empirical results show that the joint-framework achieves significant improvement over the pipeline-based baseline. Speech enhancement and speaker identity uncertainties are also shown to contribute to the system, respectively. 
In the meanwhile, multi-condition experiments suggest that our system can preserve comparable to or better results than the pipeline-based baseline for both clean and noise scenarios.


\begin{figure*}[htb]
    \vspace{-4ex}
\begin{minipage}[b]{1.0\linewidth}
  \centering
  \centerline{\includegraphics[width=18cm]{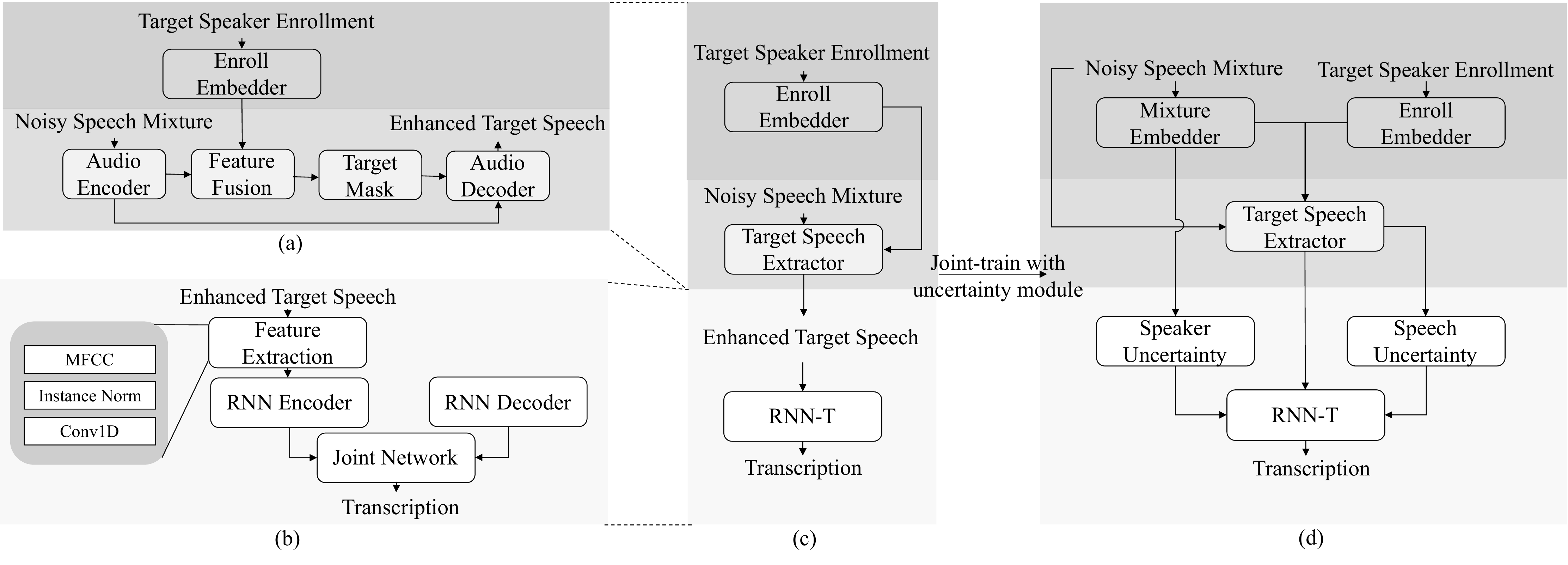}}
\end{minipage}
    \vspace{-4ex}
\caption{Architectures in this Paper: (a) is the target-speaker speech recognition model discussed in Section \ref{ssec: speech extraction}; (b) is RNN-T recognizer introduced in Section \ref{ssec: rnn-t}; (c) is a joint-network of (a) and (b). Its training strategy is shown in Section \ref{ssec: multi-stage}; (d) is the joint-model with speaker identity and speech enhancement uncertainties, which is introduced in Section \ref{sec: uncertainty estimation}.}
\label{fig: overall}
    \vspace{-4ex}
\end{figure*}

\section{Target-Speaker Speech Recognition}
\label{sec: target-speaker speech recognition}

\subsection{Time-domain Target-Speaker Speech Extraction}
\label{ssec: speech extraction}
As shown in the Fig. \ref{fig: overall}a, we follow the architecture in \cite{ji2020speaker} for time-domain target-speaker speech extraction. 
The system first extracts target speaker embedding $\vec{e}_{\text{target}}$ (i.e., a fixed-dimension vector) from enrollment speech $\vec{s}_{\text{enroll}}$. 
\begin{equation}
\label{eq: sv process}
    \vec{e}_{\text{target}} = \text{EnrollEmbedder}(\vec{s}_{\text{enroll}})
\end{equation}
The enroll embedder is a Time-Delay Neural Network (TDNN). The training of the embedder adopts multi-task objective regarding $\vec{e}_{\text{target}}$, and they are combined with scaling factors $\alpha$ and $\beta$ as follows:
\begin{equation}
\label{eq: sv loss}
    \mathcal{L}_{\text{spk}} = \mathcal{L}_{\text{triplet}}(\vec{e}_{\text{target}}) + \alpha \cdot \mathcal{L}_{\text{lmc}}(\vec{e}_{\text{target}}) + \beta \cdot  \mathcal{L}_{\text{r}}(\vec{e}_{\text{target}})
\end{equation}
where $\mathcal{L}_{\text{triplet}}$ is the triplet loss~\cite{zhang2017end,zhang2018text}, $\mathcal{L}_{\text{lmc}}$ is the large margin cosine loss, and $\mathcal{L}_{\text{r}}$ is the $L_2$ regularization term .
For the next step, the extractor produces the target speech $\hat{\vec{s}}_{\text{target}}$ from speech mixture $\vec{s}_{\text{mix}}$ conditioned on $\vec{e}_{\text{target}}$ obtained from Eq.~(\ref{eq: sv process}), as follows:
\begin{equation}
\label{eq: se process}
    \hat{\vec{s}}_{\text{target}} = \text{Extractor}(\vec{s}_{\text{mix}} | \vec{e}_{\text{target}})
\end{equation}
The architecture of the extractor, shown in Fig. \ref{fig: overall}a, adopts from the Conv-TasNet \cite{luo2019conv} consisting of an audio encoder, a feature fusion module, a prediction module for target mask, and an audio decoder to reconstruct the signal. 
The encoded information from the audio mixture and speaker embedding $\vec{e}_{\text{target}}$ are fused in a feature fusion layer. 
The objective of the extractor is the Scale-Invariant Signal-to-Noise Ratio (SI-SNR) that computes between enhanced speech $\hat{\vec{s}}_{\text{target}}$ and ground truth target speech $\vec{s}_{\text{target}}$.

\subsection{Recurrent Neural Network Transducer}
\label{ssec: rnn-t}
End-to-end ASR models have received considerable interests in recent years \cite{chiu2018state, pham2019very, karita2019comparative}. 
These models have a relatively small system footprint than HMM-based systems, which allows it to be easier adapted to on-device scenarios. However, one drawback of these end-to-end architectures is that they mostly attend to whole utterances for decoding. Thus, they cannot work for real-time streaming recognition. The end-to-end RNN-T \cite{graves2012sequence}, on the other hand, has shown its effectiveness for real-time decoding \cite{rao2017exploring}. To keep the streaming feature possible in future deployment, we construct our system on RNN-T.

Unlike other joint-framework for robust ASR, our system accepts the time-domain audio signal instead of time-frequency (TF) domain features. As shown in Fig. \ref{fig: overall}b, we add a wrapper network. It first converts the input into the Mel-Frequency Cepstrum Coefficients (MFCC) and then aggregates the features with a fixed context using convolution operation. For our target speaker recognition, we first feed target speech $\hat{\vec{s}}_{\text{target}}$ into the warpper layer ``FE$(\cdot)$" to extract features and then input the features to RNN-T to predict the transcription $\hat{W}$
\begin{equation}
\label{eq: rnn-t process}
    \hat{W} = \text{RNN-T}(\text{FE}(\hat{\vec{s}}_{\text{target}}))
\end{equation}
Note that the warpper layer ``FE$(\cdot)$" is differentiable and we can perform the back-propagation algorithm for all modules jointly.

\subsection{Multi-stage Training Strategy}
\label{ssec: multi-stage}
As shown in Fig. \ref{fig: overall}c, the framework of our target-speaker speech recognition systems cascadingly consists of three main components: an enrollment embedder, target speech extraction, and speech recognizer modules. 
Given the complexity of the framework, our pilot study indicated that it is difficult to directly train a joint-network that combines all the modules.
Therefore, a multi-stage training strategy is proposed to stabilize this joint-training.

In the first stage, the enrollment embedder is trained on a speaker verification task as Eq.~(\ref{eq: sv process}). 
Then, it is jointly trained with the target speech extractor in Eq.~(\ref{eq: se process}). 
The extractor's loss follows the objective function defined in \cite{ji2020speaker}, which is an interpolation of the speaker-related loss $\mathcal{L}_{\text{spk}}$ and SI-SNR loss. In parallel, an RNN-T recognizer shown in Fig.~\ref{fig: overall}b and Eq.~(\ref{eq: rnn-t process}) is pre-trained on a corpus with only clean speech. As shown in Fig. \ref{fig: overall}c, the two pre-trained systems are combined together. In the first, we freeze the parameter sets in target speech extraction networks and fine-tune the pre-trained RNN-T with only RNN-T objective. After the above system converges, we joint-train the whole framework with all three modules. The final stage is trained with the multi-task objective defined as:
\begin{equation}
\label{eq: joint loss}
    \mathcal{L}_{\text{joint}} = \mathcal{L}_{\text{RNN-T}}(\hat{W}, W) + \gamma (\mathcal{L}_{\text{SI-SNR}}(\hat{\vec{s}}_{\text{target}}, \vec{s}_{\text{target}}) + \phi  \mathcal{L}_{\text{spk}}),
\end{equation}
where $\phi$ and $\gamma$ are weight hyper-parameters. $\mathcal{L}_{\text{spk}}$ is inherited from the speaker verification task to stabilize the speaker embedding extraction. It is a simplified version from Eq. (\ref{eq: sv loss}) that only contains $\mathcal{L}_{\text{triplet}}$ or $\mathcal{L}_{\text{lmc}}$\footnote{ Our empirical tests showed that the two losses have a similar effect to the model.}.

\section{Neural Uncertainty Estimation}
\label{sec: uncertainty estimation}

Speech enhancement and separation modules are leveraged to mitigate the speech distortion due to the noise and interfering speakers. 
However, the enhanced signals are usually not ideal compensation, since  they often contain residual noises and artifacts introduced from the algorithms. 
In conventional statistical models like Gaussian Mixture Models (GMM), the statistical formulation facilitates the development of probabilistic uncertainty-of-observation (UoO) methods \cite{deng2005dynamic, tran2014fusion}. It considers the enhanced speech predictions as random variables rather than deterministic estimations. 
The following works extended the uncertainty processing to DNNs \cite{abdelaziz2015uncertainty, nathwani2017extended, nathwani2017consistent}.
Rather than using the uncertainty propagation, this work focuses on a more straightforward way that encodes uncertainty information as additional features to the following RNN-T speech recognizer.

\subsection{Speaker Identity Uncertainty}
\label{ssec: spk uncertainty}
Even though the target speech extraction module suppresses speech from interfering speaker, the enhanced speech often contains residual noises, especially from \textit{overlapped} speech \cite{vzmolikova2019speakerbeam, wang2019voicefilter}. 
This indicates that the recognizer should process overlap and non-overlapped speech differently. 
To explicitly inform the distinct information between overlap and non-overlapped speech, we propose to use an entropy computed from a speaker identification distribution entropy as additional features.

We first introduce a mixture embedder for noisy mixtures. The network is a convolutional neural networks (CNN) with a linear projection layer. The input of the mixture embedder is Short Time Fourier Transform (STFT) frames of $\vec{s}_{\text{mix}}$. The output of the linear projection layer is frame-wise mixture embeddings $\vec{e}_{\text{mix}}(t)$ as follows
\begin{equation}
\label{eq: sid process}
    \vec{e}_{\text{mix}}(t) = \text{MixtureEmbedder}(\text{STFT}(\vec{s}_{\text{mix}})[t])
\end{equation}
where $t$ is the frame index. A further linear layer and a softmax layer then convert $\vec{e}_{\text{mix}}(t)$ into each speaker's probability. The speaker identity uncertainty at frame $t$ is defined using the entropy as follows:
\begin{equation}
\label{eq: spk entropy}
\text{U}_{\text{spk}}(t) = -\sum_{k \in \mathcal{S}}p(k, t) \cdot \log p(k,t)
\end{equation}
where $k$ stands for a speaker ID, $\mathcal{S}$ is a set of speaker IDs in the training corpus. 
To explicitly train the mixture embedder, we employ a cross-entropy loss and interpolate it when training the target speech extraction model shown in Eq.~(\ref{eq: se process}). Inspired by \cite{koizumi2020speech}, we also use the mixture embedding $\vec{e}_{\text{mix}}(t)$ as a speaker embedding of the mixture. After mean-pooling, the embedding is fed into the feature fusion module discussed in Section \ref{ssec: speech extraction} with target speaker embedding.

\subsection{Speech Enhancement Uncertainty}
\label{ssec: speech uncertainty}
Three uncertainty estimators are frequently used in uncertainty estimation for enhanced speech, including oracle uncertainty (OU) estimator \cite{tran2015nonparametric}, Kolossa's uncertainty (KU) estimator \cite{abdelaziz2015uncertainty, nathwani2017consistent}, and Neural network-based uncertainty (NNU) estimator \cite{nathwani2017extended, tran2015nonparametric}. The OU estimator is the best possible uncertainty estimator. The OU for frame $t$ is defined as follows:
\begin{equation}
\label{eq: OU}
\text{OU}(t) = (\mathbf{y}(t) - \mathbf{\hat{y}}(t))^2
\end{equation}
where $\mathbf{y}$ and $\mathbf{\hat{y}}(t)$ stands for clean and enhanced speech features (e.g., MFCC of $\vec{s}_{\text{target}}$ and $\hat{\vec{s}}_{\text{target}}$), respectively. As the OU estimator needs information of clean speech, it is not available for practical usage. NNU estimators are the state-of-the-art methods that train neural networks to predict the OU. As shown in \cite{nathwani2017extended}, they achieve comparable performance to the OU estimator.

NNU estimators in previous works were often using DNN architectures and were trained separately using noisy \& enhanced speech features (e.g., MFCC) as inputs and OU as targets \cite{nathwani2017extended, tran2015nonparametric}. In this work, we extend the NNU estimator to a convolutional neural uncertainty (CNU) estimator that jointly trains the NNU estimator with the target-speaker speech recognition system discussed in Section \ref{ssec: multi-stage}. 
The CNU contains an InstanceNorm layer, a convolutional layer, a linear layer with the Leaky-ReLU activation function, and a prediction layer. 
The CNU is applied to the system as follows:
\begin{equation}
\label{eq: CNU proces}
\hat{\text{OU}}_{\text{norm}} = \text{CNU}(\mathbf{\hat{y}}, (\mathbf{y}_{\text{mix}} - \hat{\mathbf{y}})^2)
\end{equation}
where $\mathbf{y}_{\text{mix}}$ is the mixture speech features, the inputs of the CNU are enhanced speech features $\mathbf{\hat{y}}$ and a distance $(\mathbf{y}_{\text{mix}} - \hat{\mathbf{y}})^2$, the prediction target is the $\text{OU}_{\text{norm}}$ .\footnote{We use InstanceNorm for OU normalization.} We adopt L1 loss as the training objective of CNU. We use the hidden states of the linear layer as speech enhancement uncertainty features $\text{U}_{\text{speech}}$ and concatenate it with other recognizer features. Our empirical studies showed that using the hidden states instead of the $\hat{\text{OU}}_{\text{norm}}$, is more stable in training.

\subsection{Proposed System with Uncertainty}

Our proposed system with uncertainty training is shown in Fig. \ref{fig: overall}c, including three steps. In the first step, the system extracts a mixture embedding and an enrolled embedding. The mixture embedding is extracted from noisy speech mixture using a mixture embedder discussed in Section \ref{ssec: spk uncertainty}. The enrolled embedding is from enrollment speech shown in Section \ref{ssec: speech extraction}. Then, the two embeddings' concatenation is fed into the target speech extractor with a noisy speech mixture for the second step. The output of this step is the enhanced time-domain speech signal. For the last step, we extract speaker identity uncertainty using the mixture embedder's probabilistic distribution in \ref{ssec: spk uncertainty} and speech enhancement uncertainty features using CNU proposed in \ref{ssec: speech uncertainty}. We then concatenate the uncertainty measures with the frequency features (i.e., MFCC) of the enhanced speech and feed it into RNN-T.

The model is also trained in a multi-stage manner. First, the enroll embedder and RNN-T for clean speech recognition are trained separately. Next, they are subsumed into the target-speaker speech extraction for joint-training.\footnote{The mixture embedder also trains at this step.} At last, we freeze the target speech extraction network and fine-tune the pre-trained RNN-T system but with two additional uncertainty features. Noted that the CNU for $\text{U}_{\text{speech}}$ is also trained in the last stage. The objective at this stage is with multi-task learning as follows:
\begin{equation}
\label{eq: joint loss2}
    \mathcal{L}_{\text{joint2}} = \mathcal{L}_{\text{RNN-T}} + \mathcal{L}_{\text{CNU}}
\end{equation}
Our experiments found that the model with uncertainty did not show significant improvements after joint-training as a whole model. Therefore, we do not additionally join-train the whole model.

\section{Experiments}
\label{sec: exp}

\subsection{Dataset}
For the training of enroll embedder, a subset of King-ASR-216 and King-ASR-210 is chosen for pre-training. The pre-processing is the same as in \cite{ji2020speaker}. The other experiments are trained on AISHELL-2 \cite{du2018aishell} and its noisy simulation corpus. For each utterance, we mix it with utterances from other speakers at a randomly signal-to-interference ratio (SIR) in \{0dB, 6dB, 12dB\}. The number of interfering speakers in each mixture is randomly selected from \{0, 1, 2\}. Environmental noise is added to the mixture as well with signal-to-noise ratio (SIR) randomly ranged from \{6dB, 12dB, 18dB, 24dB, 30dB\}. We adopt the same noise set as \cite{ji2020speaker}, which including several ambient noises from ``daily life" environments. Different room impulse responses are considered in the simulation. All the above samplings are with equal chance. The final simulation corpus contains 963,469 utterances for training, 20,347 utterances for development, and 25,382 utterances for testing. \footnote{Noted that the speakers in each subset are not overlapped.}

\subsection{Network and Training Settings}
\label{ssec: network and training settings}
For our proposed model, the enroll embedder and the target-speaker speech extraction module adopted the same configuration as \cite{ji2020speaker}. The RNN-T system has four encoder layers and two decoder layers. Each layer is a bi-directional LSTM layer and contained 512 hidden states with a 0.2 dropout rate. The MFCC layer in Fig.~\ref{fig: overall}b uses 25ms window length, 10ms window shift, 512-point FFT, and 40 Mel bins. The Conv1D shown in Fig.~\ref{fig: overall}b has a kernel size of 7 and stride of 3 for feature down-sampling. The CNU discussed in Section \ref{ssec: speech uncertainty} adopts a Conv1D layer with 128 output channels and 11 kernel size but without stride. The following FC layer (i.e., the speech enhancement uncertainty features) has 32 dimensions. For all the stages, we adopt Adam optimizer. For most stages, the initial learning rate is 0.001, and it is annealed by half if no improvement is observed for the next epoch. However, for joint training, each component is attached with optimizers of different learning rates: the initial learning rates ware set to 1e-5 for RNN-T, 1e-7 for target speech extractor, 2e-7 for the enroll embedder. $\alpha$ and $\beta$ in Eq.~(\ref{eq: sv loss}) are set to 0.2 and 0.001. $\gamma$ and $\phi$ in Eq.~(\ref{eq: joint loss}) are set to 0.01 and 1.0.

We conduct the experiments in two folds. For the first (Noisy Training), we only train and test our model in noisy settings. We compare seven models: Model \textbf{I} that is trained on original AISHELL-2 using the model in Section \ref{ssec: rnn-t}, Model \textbf{II} that directly adopts the pre-trained Clean RNN-T and target speech extraction module, Model \textbf{III} that freezes the target speech extraction module and fine-tunes the Clean RNN-T, Model \textbf{IV} that jointly trains the whole network from model \textbf{II}, Model \textbf{V}, \textbf{VI}, \textbf{VII} fine-tuned RNN-T and froze the target speech extraction module with speaker identity ($\text{U}_{\text{spk}}$), speech enhancement ($\text{U}_{\text{speech}}$), and both ($\text{U}_{\text{both}}$) uncertainty  measures, respectively. For the second fold (Multi-condition Training), we present the stability of the system with multi-condition (i.e., clean and noise) training. Four models including \textbf{III}, \textbf{V}, \textbf{VI}, \textbf{VII} are trained while Model \textbf{I} is presented as references. We employ an on-the-fly loader that randomly fed clean and noisy speech for model training. As clean utterances introduce artifacts for uncertainty training\footnote{the OU for clean utterances are zero matrices.}, we adopt 20\% training clean speech ratio for models using uncertainty, but 50\% for other models.

\subsection{Results}

The CER results are shown in Table \ref{tab: exp result}. Among the models that do not consider uncertainty features, model \textbf{IV} with joint-training reaches the best performance. As discussed in Section \ref{ssec: multi-stage}, the model is trained by the multi-stage order of ``\textbf{I}-\textbf{II}-\textbf{III}-\textbf{IV}". The results in Table \ref{tab: exp result} indicate the system can receive progressive improvements over each stage. Our best results come from the model with both speech enhancement and speaker identity uncertainty. Even though the uncertainty only trained with a frozen target speech extraction module, both model \textbf{VI} and model \textbf{VII} significantly outperform the model \textbf{IV} with joint-training. Meanwhile, the speaker identity uncertainty does not offer enough improvement to the system when it is introduced alone. It significantly reduces the CER when it is combined with the speech enhancement uncertainty. 

Table \ref{tab: multi-con exp result} shows the multi-condition experiment results. Generally speaking, the noisy test results are aligned with the noisy training experiments in Table \ref{tab: exp result}. The clean test results suggest that training using a multi-condition setting can also elevate the model performances for clean speech recognition. Note that the speech enhancement uncertainty may down-grade the clean speech recognition performance (see model \textbf{VI} in Table \ref{tab: multi-con exp result}) because of the artifacts discussed in Section \ref{ssec: network and training settings}. However, the speaker identity uncertainty has shown its effectiveness in eliminating the instability, as shown in model \textbf{VII} in Table \ref{tab: multi-con exp result}.

\begin{table}[]
    \vspace{-1ex}
\centering
\caption{\label{tab: exp result} Noisy Training Result (CER): Model \textbf{I} is a RNN-T trained on clean speech, others were trained and tested on noisy data only; Model \textbf{II} is a pipeline system that used pre-trained modules; $\star$ means that we fine-tuned RNN-T while froze the pre-trained target speech extraction module; $\dagger$ stands for the model using joint-training. Model \textbf{V}, \textbf{VI}, \textbf{VII} employed speaker identity uncertainty ($\text{U}_{\text{spk}}$), speech enhancement uncertainty, ($\text{U}_{\text{speech}}$), and both uncertainties ($\text{U}_{\text{both}}$), respectively.}

\begin{adjustbox}{max width=0.98\linewidth}
     
\begin{tabular}{llllll}

\hline
\textbf{Class} & \textbf{Model} & \textbf{Overall} & \textbf{1Spk} & \textbf{2Spk} &\textbf{3Spk} \\
\hline
\multirow{3}{*}{\textbf{Baseline}} & \textbf{I}  & 81.9 & 20.4 & 84.0 & 139.0   \\
& \textbf{II}  & 45.9 & 13.1 & 37.2 & 85.7  \\

 & \textbf{III}$^\star$  & 26.2 & 12.5 & 26.2 & 39.3   \\
 \hline
\multirow{1}{*}{\textbf{Joint-Train}} & \textbf{IV}$^\dagger$  & 24.4 & 11.1 & 24.5 & 37.1 \\
\hline
\multirow{3}{*}{\textbf{Uncertainty}} & \textbf{V}$^\star$ ($\text{U}_{\text{spk}})$ & 25.8 & 12.2 & 25.1 & 39.5  \\
& \textbf{VI}$^\star$ ($\text{U}_{\text{speech}}$)  & 22.5 & 10.0 & 22.0 & 34.9  \\
& \textbf{VII}$^\star$ ($\text{U}_{\text{both}}$)  & \textbf{21.8} & \textbf{9.6} & \textbf{21.3} & \textbf{33.8}  \\
\hline
\end{tabular}
\end{adjustbox}

    \vspace{-3ex}

\end{table}

\begin{table}[]
\caption{\label{tab: multi-con exp result} Multi-condition Results (CER): Model \textbf{I} is a RNN-T trained on clean speech, others were trained on a combination of noisy and clean data; $\star$ means that we fine-tuned RNN-T while froze the pre-trained target speech extraction module; Model \textbf{V}, \textbf{VI}, \textbf{VII} employed speaker identity uncertainty ($\text{U}_{\text{spk}}$), speech enhancement uncertainty, ($\text{U}_{\text{speech}}$), and both uncertainties ($\text{U}_{\text{both}}$), respectively.
}
\begin{adjustbox}{max width=0.98\linewidth}
\centering
\begin{tabular}{lclll}
\hline
\textbf{Model} & \textbf{All (Clean/Noisy)} & \textbf{1Spk} & \textbf{2Spk} &\textbf{3Spk} \\
\hline
\textbf{I} & 10.3/81.9 & 20.4 & 84.0 & 139.0   \\
\textbf{III}$^\star$ & \textbf{9.9}/23.7 & 9.8 & 23.8 & 37.0   \\
\hline
\textbf{V}$^\star$ ($\text{U}_{\text{spk}}$) &  11.3/23.5 & 10.0  & 23.1  &  36.9  \\
\textbf{VI}$^\star$ ($\text{U}_{\text{speech}}$) & 11.7/22.5 & 9.1 & 21.5 & 36.4  \\
\textbf{VII}$^\star$ ($\text{U}_{\text{both}}$) & 10.0/\textbf{21.5} & \textbf{9.0} & \textbf{21.2} & \textbf{33.8}  \\
\hline
\end{tabular}
\end{adjustbox}
    \vspace{-3ex}

\end{table}

The uncertainty features are especially useful when handling multi-speaker speech. Our further investigation found that the features can significantly reduce the recognition insertion error for the recognizer. As shown in Table \ref{tab: i&d&s errors}, Model \textbf{VII} gained 17.5\% relative insertion error reduction comparing to Model \textbf{III} for the three speaker case in noisy training (i.e., Table \ref{tab: exp result}). For multi-condition training (i.e., Table \ref{tab: multi-con exp result}), Model \textbf{VII} received similar 15.5\% relative insertion error reduction comparing to Model \textbf{III}. The insertion errors are likely to happen when there are residual noises from other speakers in the enhanced speech. Thus, the reduction of the insertion error rate aligns with our motivation of adding uncertainty features.

\begin{table}[htb]
    \vspace{-3ex}
    \centering
        \caption{Error Distribution Between Models with and without Uncertainty. The results are from three speaker cases in noisy training.}
        \begin{adjustbox}{max width=0.98\linewidth}
    \begin{tabular}{lllll}
\hline
\textbf{Model} & \textbf{Insertion} & \textbf{Deletion} & \textbf{Substitution} \\
\hline
\textbf{III} & 2123 & 3321 & 16075 \\
\textbf{VII} & \textbf{1751} & \textbf{3260} & \textbf{13673} \\
\hline
    \end{tabular}
    \end{adjustbox}
        \vspace{-4ex}

    \label{tab: i&d&s errors}

\end{table}

\section{Conclusion}
\label{sec: conclusion}
This work presents a joint-framework combining target speech extraction and RNN-T for target-speaker speech recognition. We explored the framework with a multi-stage training strategy to stabilize the training. Next, we designed two novel uncertainty measures and subsumed them into the system. Specifically, we propose the speaker identity uncertainty based on entropy and the speech enhancement uncertainty based on a neural uncertainty estimator. Our experiments on simulated AISHELL-2 corpus show that both the joint-train framework and the uncertainties significantly improve the system performance.



\bibliographystyle{IEEEbib}
\bibliography{strings,refs}

\end{document}